# Design of High-Quality Reflectors for Vertical Nanowire Lasers on Si

Xing Zhang[‡,∥,†], Hui Yang[#,†], Yunyan Zhang[‡,§,*], Huiyun Liu[‡]

[‡] Department of Electronic and Electrical Engineering, University College London, London WC1E 7JE, United Kingdom

[#] Department of Materials, Imperial College London, Exhibition Road, London SW7 2AZ, United Kingdom

[§] Department of Physics, Universität Paderborn, Warburger Straße 100, 33098, Paderborn, Germany

[∥] Faculty of Electrical Engineering and Computer Science, Ningbo University, Ningbo, 315211, China

[†] These authors contributed equally to this work.

*E-mail: yunyan.zhang.11@ucl.ac.uk, yunyan.zhang@uni-paderborn.de

**ABSTRACT**: Nanowires (NWs) with a unique one-dimensional structure can monolithically integrate high-quality III-V semiconductors onto Si platform, which is highly promising to build lasers for Si photonics. However, the lasing from vertically-standing NWs on silicon is much more difficult to achieve compared with NWs broken off from substrates, causing significant challenges in the integration. Here, the challenge of achieving vertically-standing NW lasers is systematically analyzed. The poor optical reflectivity at the NW/Si interface results severe optical field leakage to the substrate, and the commonly used $SiO_2$ or $Si_2N_3$ dielectric mask at the interface can only improve it to ~10%, which is the major obstacle for achieving low-threshold lasing. A NW super lattice distributed Bragg reflector is therefore proposed, which is able to greatly improve the reflectivity to >97%. This study provides a highly-feasible method to greatly improve the performance of vertically-standing NW lasers, which can boost the rapid development of Si photonics.

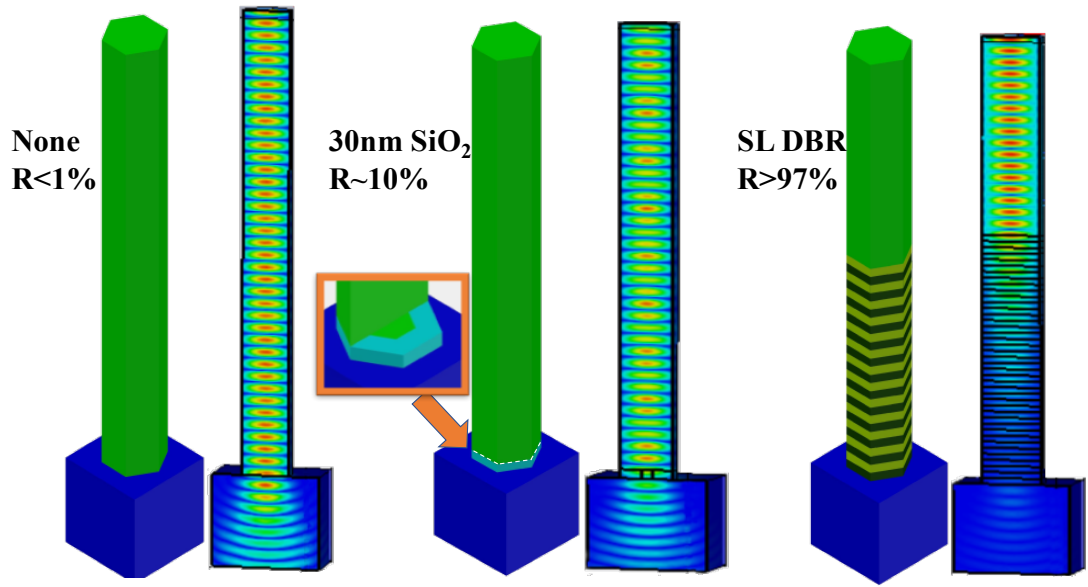

Semiconductor nanowires (NWs) lasers have a strong optical mode confinement in a very small volume, allowing lasing with diameters of a few hundred nanometers and length of a few microns.[1-3] Their small-size, low-power features endow them a wide range of novel applications,[4] including Si photonics that with a big market.[5] Si photonics using optical signals offers huge advantages in speed and consumption-energy saving compared with currently-used copper-wire technology using electrical signals.[6,7] However, Si has an indirect band gap and hence low light emitting efficiency. The lack of high-quality low-cost lasers as the signal sources on Si chips is the bottleneck of this technology. III-V materials have a direct-bandgap and therefore good optical properties, which are ideal for making high-efficiency lasers. However, when growing III-V thin-film components onto silicon, it is difficult to balance crystal quality, uniformity and area of selective growth, due to the large lattice, thermal properties, and atomic polarity differences.[8-10] When in the form of NWs, high-quality III-V materials can be monolithically grown on a Si platform.[11-13] This enables the fabrication of high-quality photonics light sources for low-cost, ultra-high density integration on Si, solving one of the major challenges that have been limiting the Si photonics.[14-16,18]

Whilst optically-pumped NW lasers have been widely demonstrated,[19-23] majority of them were broken off from their original substrate and transferred to holder substrates. In this structure, both of the NW end facets are exposed to air with a low refractive index (n=~1). The large refractive-index difference at the interface is beneficial for reflecting light back into the resonant cavity and form inverted carrier distribution. Despite the simplicity of achieving lasing with horizontal structure, NW lasers in vertically-standing structure are more desirable in most occasions due to the advantages, including massive and large-area integration, accurate allocation of NW lasers on the Si chip, simplified fabrication processes, and suitable for 3-D circuit design.

However, to achieve lasing with vertically-standing structure is much more challenging, and there are only few researches on optically-pumped vertically-standing NW lasers so far.[24-26] As it has been reported, the internal quantum efficiency of the vertical NW lasers is only 16.8%.[25] This can cause enormous challenges for the fabrication of electrically-pumped lasers. One possible reason could be the similar refractive-index difference between the Si substrates and III-V materials (*e.g.* $n_{Si}/n_{GaAs}$=~3.63/~3.59 at 870 nm) that can cause sever optical-field leakage from the NW into the substrate and greatly degrade the quality of optical resonant cavity. Therefore, the detailed knowledge on the causes and how to further improve performance of NW lasers are urgently needed.

In traditional film-film devices, the super lattice (SL) distributed Bragg reflectors (DBRs) are widely used to provide strong light reflection.[27,28] To build high-quality DBR, material combinations with large refractive-index difference is preferred. In the traditional thin-film growth, the strict requirement of small strain between epi-layers and substrates excludes a significant material combinations that can provide large refractive-index differences.[29] NWs with a unique one-dimensional structure and a small cross-section can provide very efficient lateral and axial strain relaxation, which can offer more advantages in this structure construction and allow various material combinations that are not allowed in thin-film structures. There have been reports on defect-free heterojunctions in NWs with large lattice mismatches, e.g. InAs/InP, InAs/InSb, GaAs/GaP.[30-32] Besides, the use of III-V-V type of materials can produce heterojunctions with atomic sharpness, which is beneficial for achieving abrupt refractive-index change.[30] All these advantages can be used to construct high-quality

SL DBR inside NWs, which is however no report is given on this topic so far as we know.

In this work, the light propagation and transmission behavior in NWs vertically standing on Si has been systematically studied by numerical modeling. The sever light leakage is found from NWs to the Si substrate even with the presence of $SiO_2$ and $Si_2N_3$. A SL DBRs structure is proposed, which can effectively solve this issue.

The software used here is Comsol Multiphysics based on FEM (Finite Elements Method). Based on the RF module integrated in this software, the models of NW lasers were built based on the observation from our experiments,[33,34] and the inner light transmission field inside NWs was simulated. This research started with GaAs(P) NW lasers, because majority of reports in literature are GaAs-based and the obtained knowledge can also be used to other material systems. The NW laser structure can be seen in Figure 1a. The GaAs NW is standing vertically on a Si substrate with a diameter of ~500 nm and length of ~10 um, acting as both the gain material and the resonant cavity. The center wavelength of laser is selected to be 870 nm which is the GaAs emission wavelength at the room temperature. We assume the laser beams with the same intensity propagating from the NW to the NW/Si interface and integrate the reflection energy right above the interface.

GaAs NWs are commonly grown directly on Si substrates without special interface structure to enhance the light reflection, which is shown in Figure 1a. As can be seen from the light-field distribution shown in Figure 1b, there is severe light energy leakage from the NW into the substrate. The refraction-index difference at the NW/Si interface is very small for the wavelength >500nm shown in Figure 1c. As a result, the reflection at the interface is ~$4\times10^{-5}$ at the wavelength 850~890nm (Figure 1d). This phenomenon is highly beneficial for constructing tandem photovoltaics, such as the III-V NW/Si two junction solar cells. As demonstrated by Diedenhofen et al,[35] more than 90% of the photons, with the energy lower than the InP absorption, were coupled into the underlying substrate by the InP NWs. However, it is hardly form any proper resonant

Page 4 of 15

cavity for lasers with this configuration. Therefore, proper structure design is needed to reduce light leaking into the substrate.

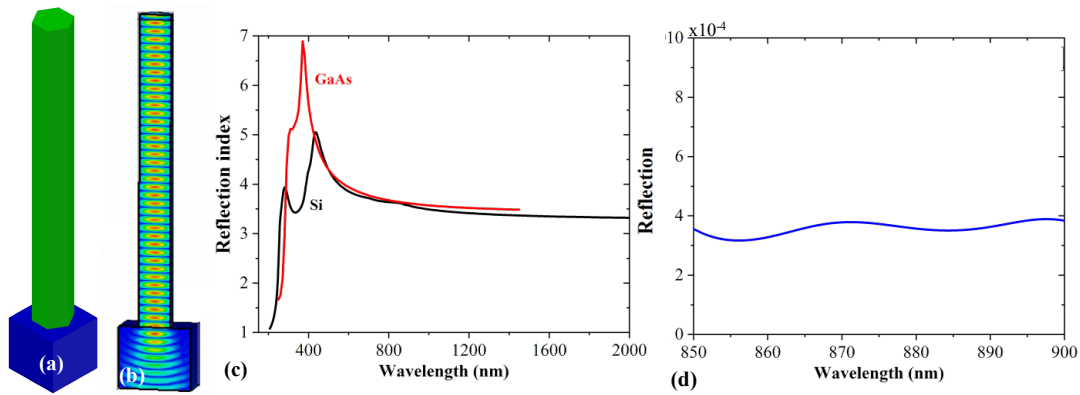

Figure 1. The light propagation and transmission behavior of a GaAs NW standing directly on Si without any special interface structure to enhance the light reflection. (a) Illustration of the structure. (b) light-field distribution at 870nm of (a). (c) Wavelength-dependent refraction-index difference between GaAs and Si. (d) Wavelength-dependent reflection at the NW/Si interface.

$Si_3N_4$ (n=~2.347 at 870 nm) and $SiO_2$ (n=~1.44 at 870 nm) are widely used as the pattern mask to do the positioned NW growth on Si substrate,[36] and they have a much lower refractive index compared with III-Vs (e.g. $n_{GaAs}$=~3.5 at 870 nm). Thus, they can be potentially used to enhance the light reflection at the interface. A ring-shaped dielectric reflector is therefore studied here (Figure 2a). The ring hole in the centre of the reflector is the opening for the NW growth with substrate lattice template. As can be seen from the light-field distribution in the NWs in Figure 2b, the optical field leakage into the substrate reduces compared with Figure 1b when the dielectric layer with a thickness of 100nm is introduced at the interface. Both dielectric layers with a large ring width, defined to be the thickness of concentric hexagon with the outer edge the same size as the NW, can increase the reflection to above 30%, which is shown in Figure 2c. However, the interface reflectance is small (<3%) for small ring widths and rises slowly with the increase of the width; only when the width is approaching the NW diameter, the reflectance rises rapidly. This phenomenon is the same for NWs of different diameters we study (Figure 2d), which suggest that the width of dielectric needs to be at least over half the size of the NW to get more apparent reflection. It needs to be noted that the interface reflection of NWs with a $SiO_2$ layer can be up to 67% higher than that of $Si_3N_4$ layer at a large ring width, which is due to the larger index difference with the NW. Consequently, it would be better to choose $SiO_2$ as the reflector material when building lasers. Moreover, the thickness of the dielectric layer can greatly influence the reflectance. As is shown in Figure 2e, the reflection of the NWs with a $SiO_2$ dielectric layer grows almost linear to ~50% at 100 nm. At present, the thickness of dielectric pattern is normally between 20-30 nm, which can only provide a low reflectivity of ~10%. We have not notice the reports using thicker patterns. More study is needed about growing on super-thick patterns, such as 100 or 200 nm, which can however bring high challenges on both the pattern making and NW growth.

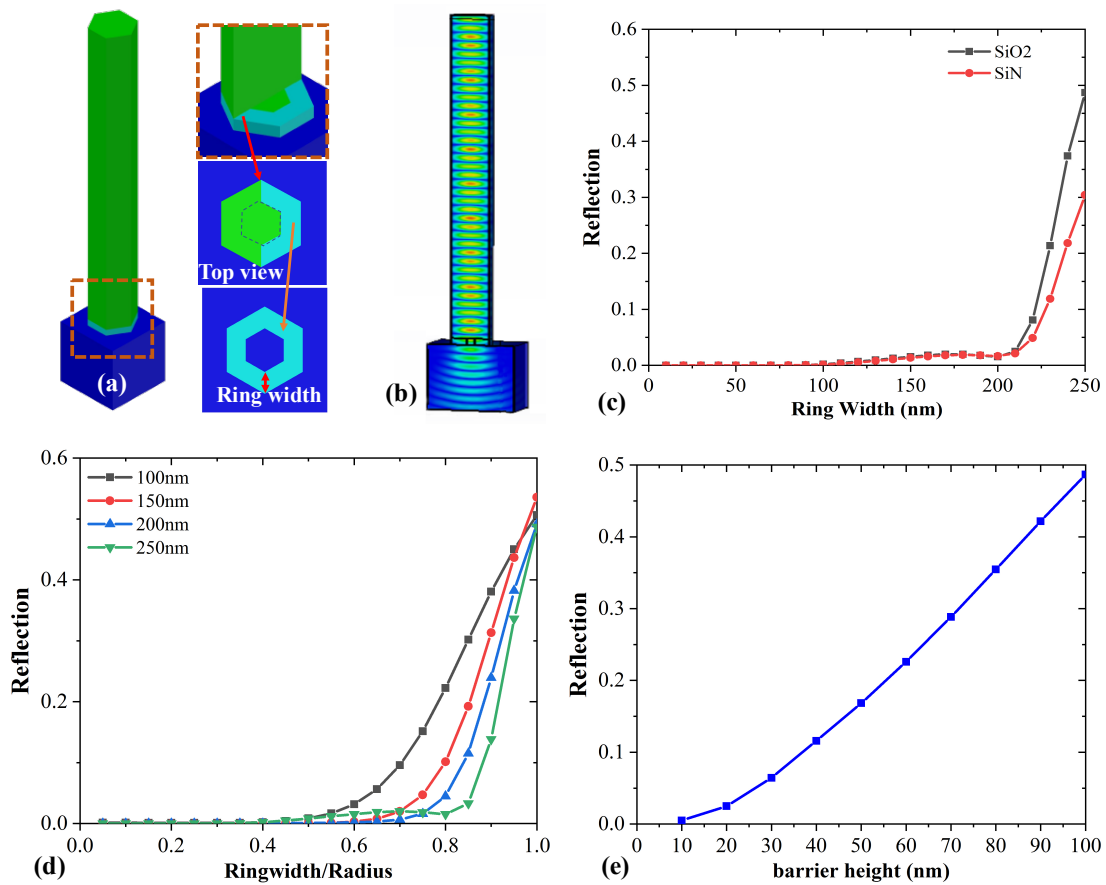

Figure 2. The light propagation and transmission behaviour of a GaAs NW (500 nm) standing directly on Si with a hexagon ring dielectric layer to enhance the light reflection. (a) Illustration of the dielectric ring reflector. (b) light-field distribution at 870nm of the GaAs NW with a 230nm-width 100nm-thick SiO$_2$ interface layer. (c) Ring-width-dependent reflection of a GaAs NW with a 100 nm-thick of SiO$_2$ or Si$_3$N$_4$ layer. (d) Influence of ring-width/radius ratio on the reflection of NW of different size. (e) Dielectric thickness-dependent reflection of a GaAs NW with a 250-nm width SiO$_2$ interface layer.

With the assumption that a 100-nm-thick dielectric reflector between the NW/Si interface can be easily used, the reflection can only be improved to ~50% and a high percentage of the light is still lost through the substrate, causing high threshold of electrically-pumped lasers. Considering the advantages of NWs in the construction of III-V-V heterojunctions as mentioned above, a GaAs$_{0.1}$P$_{0.9}$/GaAs ($n_{GaAsP}$/$n_{GaAs}$=3.2/3.59) SL DBR can be introduced between NW and Si substrate (Figure 3a). As can be seen from the light-field distribution in Figure 3b, most of light field is resonating inside the NW with only little light energy leaks into the substrate. The SL period is one of the key parameters that decide the peak reflection wavelength. For a lasing wavelength of 870nm, a high reflectance of over 80% can be achieved with a period between 124nm to 142nm, and the reflection reaches the maximum can be 96% at 136nm (Figure 3c). The pair of SL can also significantly influence the light reflection. As can be seem in Figure 3d, the reflection increases rapidly with the increase of the SL pairs to 15 and gradually saturates after 20 pairs. The reflection at 870 nm can reach 0.96 with a 20-pair SL and 0.98 with a 30-pair SL. It needs to be mentioned that the SL DBR can have a large reflection window. For example, its reflectance can maintain above 90% between 820nm to 880nm. Thus, the SL DBR has significant advantages over the dialectic reflectors.

The technique is well developed for growing the SL DBR on substrate surface in the form of thin-film structure. Integrating NWs directly onto the substrates with DBRs (Figure 3f) can provide a possibility to avoid the complexity of growing DBR into NWs, which has been widely used experimentally.[37-39] Thus, the influence of the location on the refection effect of SL DBR is also studied with the optimized DBR parameters used in Figure3c. As can be seen from Figure 3g, the DBR can effectively avoid the leakage of the optical field into the substrate. However, the field disperses after coming out of the NW bottom into the substrates, and a significant portion of the energy can no longer be reflected back to the NWs with a small cross section. As supported by the reflection spectrum in Figure3h, the energy in the NW at the peak wavelength can only maintain at ~60%, which is greatly reduced compared with that of the NW DBR.

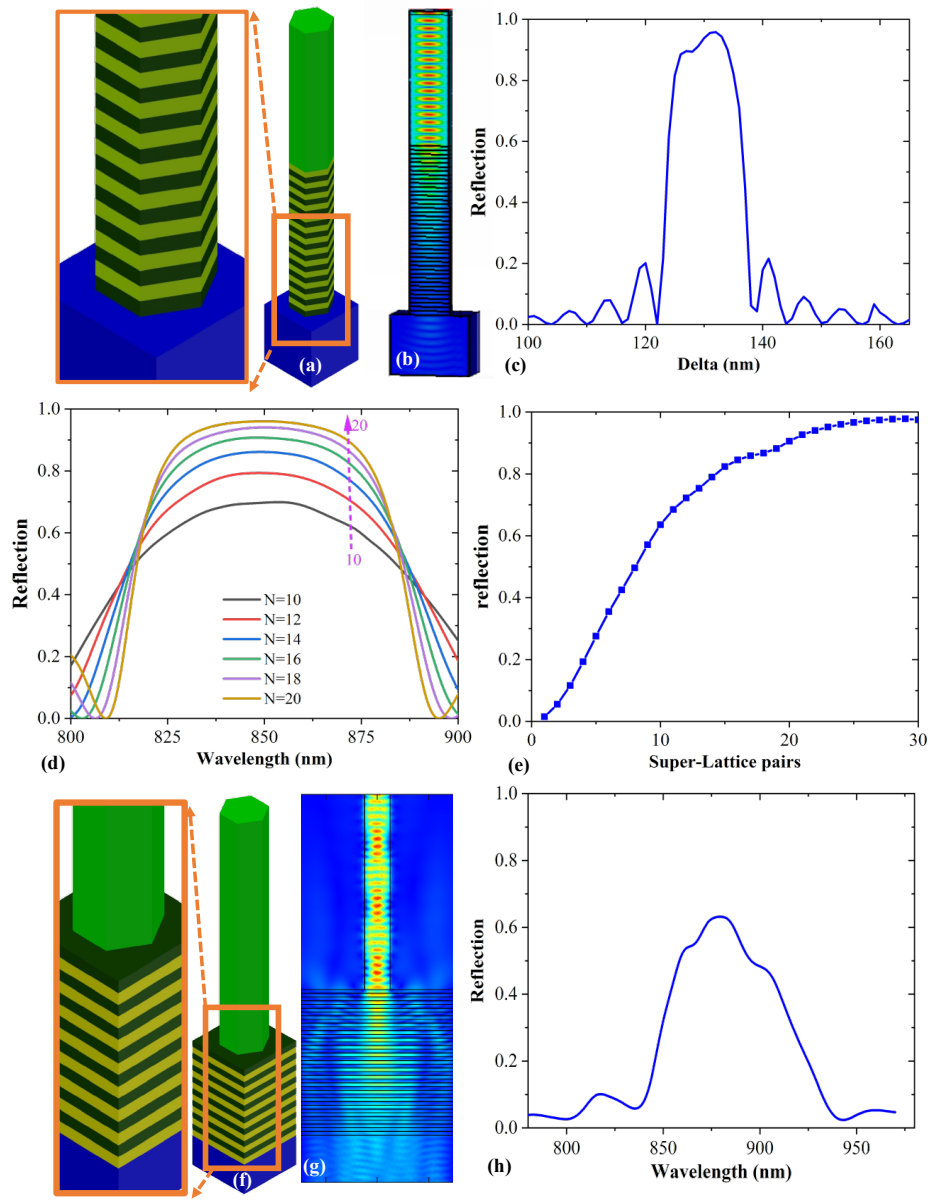

Figure 3. The light propagation and transmission behavior of a GaAs NW standing directly on Si with a SL DBR to enhance the light reflection. (a)~(e) are from the structure with a NW DBR reflector. (a) Illustration of the structure. (b) light-field distribution at 870nm of the NW. (c) SL-period-dependent light reflection at 870 nm. (d) Reflection spectra of DBRs with different SL pairs. (e) SL-pairs-dependent light reflection at 870 nm. (f)~(h) are from the structure with a substrate DBR reflector. (f) Illustration of the structure. (g) light-field distribution at 870nm of the NW. (h) light reflection spectrum at the NW/substrate interface.

We also extend the study method to telecom communication wavelength. The NWs used here are made of InAsP. The InAs/InP lattice periods are 0.236μm for 1310nm light and 0.28μm for 1550nm light respectively. As can be seen in Figure 4, the SL DBR can also work well with these lasers and both of them can improve the light reflection to ~90%

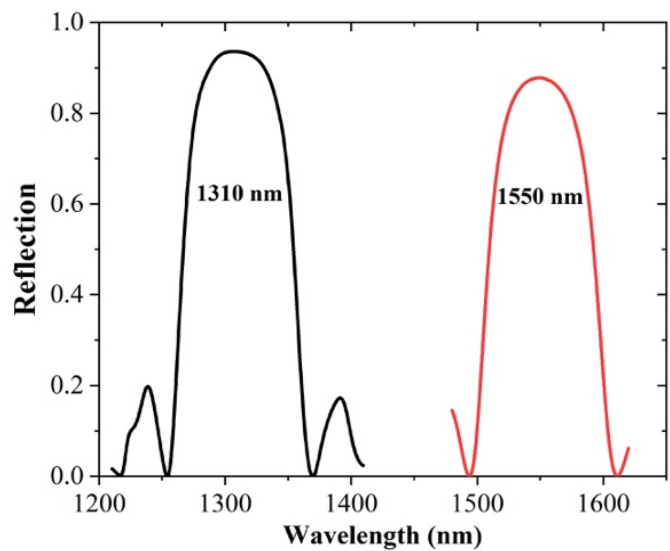

Figure 4. Reflection spectra of SL DBRs that work for 1310nm and 1550nm lasers.

In summary, the challenge of achieving vertically-standing NW lasers on Si is systematically analyzed. The optical reflectivity at the GaAs NW/Si interface is ~$4 \times 10^{-5}$ at the wavelength 850~890nm, which is due to the small refraction-index difference between GaAs and Si. This small reflectivity prevents the formation of acceptable optical resonant cavity for the realization of lasing. $SiO_2$ and $Si_2N_3$, with larger refraction-index differences with Si, are widely used in the NW growth, which is however can only improve it to ~10% and majority of the energy is still leaking into the substrates. To further improve the reflectivity at NW/substrate interfaces, the NW SL DBR is proposed, which can greatly improve the reflectance to 97%, allowing the formation of high-quality optical resonant cavity. Besides, these reflectors can also be used for telecom communication wavelength (1310nm and 1550nm) and a high reflectance of ~90% can be achieved. This research provides useful information for the design of high-quality vertically-standing NW lasers onto Si platform, which can be widely used in the integrated circuits.

**NOTES:**

The authors declare no competing financial interest.

**ACKNOWLEDGEMENTS**

The authors acknowledge the support of Leverhulme Trust, EPSRC (grant nos. EP/P000916/1, EP/P000886/1, EP/P006973/1), and EPSRC National Epitaxy Facility.